\documentclass[a4paper]{article}
\usepackage[T1]{fontenc}
\usepackage[utf8]{inputenc}
\usepackage{float}
\usepackage{bm}
\usepackage{amsmath}
\usepackage{amsthm}

\makeatletter

\newcommand*\LyXZeroWidthSpace{\hspace{0pt}}

\providecommand{\tabularnewline}{\\}

\@ifundefined{date}{}{\date{}}

\usepackage{color}
\usepackage{amsfonts}\usepackage{amsthm}\usepackage{amscd}\numberwithin{equation}{section}
\usepackage{bm}
\usepackage[all]{xy}

\usepackage{algorithm}
\usepackage{algorithmicx}
\usepackage{algpseudocode}
\usepackage[most]{tcolorbox}
\tcbuselibrary{listingsutf8}
\usepackage{lipsum}
\tcbset{
  highlight algorithm/.style={
    colback=gray!10,
    colframe=black,
    boxrule=0.5pt,
    arc=2pt,
    left=6pt,
    right=6pt,
    top=6pt,
    bottom=6pt,
    width=\textwidth,
    sharp corners,
  }
}

\usepackage{xcolor}

\usepackage[english]{babel}

\makeatother

\begin{document}
\title{Information Entropy and Dissipation Potential in Stochastic Nambu
Nonequilibrium Thermodynamics}
\author{So Katagiri$^{\dagger}$\thanks{So.Katagiri@gmail.com}, Tatsuaki
Wada$^{\diamond}$\thanks{tatsuaki.wada.to@vc.ibaraki.ac.jp}}

\maketitle
$^{\dagger}$\textit{Major in Complex Systems Science, Graduate School
of Science and Engineering, Ibaraki University, Nakanarusawa-cho,
Hitachi, 316-8511, Japan}.\\
 $^{\diamond}$\textit{Region of Electrical and Electronic Systems
Engineering, Ibaraki University, 4-12-1 Nakanarusawa-cho, Hitachi-shi
316-8511, Japan} 
\begin{abstract}
We investigate the relationship between the dissipation potential
$S_{\mathrm{NB}}$ and information entropy in stochastic Nambu nonequilibrium
thermodynamics (NNET). Starting from pure diffusion, gradient-flow
systems, and systems with Hamiltonian rotational currents, we derive
the stationary distribution of the Fokker--Planck equation including
a Nambu flow. We show that, when the Nambu flow preserves the dissipation
potential and an Einstein-type relation holds between the diffusion
matrix and the transport tensor, the stationary distribution is governed
by an exponential weight determined by $S_{\mathrm{NB}}$. Consequently,
the local self-information is directly related to the dissipation
potential, demonstrating that $S_{\mathrm{NB}}$ should not be identified
with the Shannon entropy itself but instead serves as the stationary
quasipotential governing the nonequilibrium steady-state distribution.
Using the distinction between the thermodynamic entropy and the dissipation
potential established for an open gas--piston system coupled to a
thermal bath and an external-pressure reservoir, we further clarify
the Massieu interpretation of $S_{\mathrm{NB}}$, introduce a Nambu
canonical distribution, and derive the covariance of fluctuations
within the saddle-point approximation. These results provide a statistical
interpretation of the NNET dissipation potential by identifying it
as the stationary quasipotential underlying nonequilibrium steady
states. 
\end{abstract}

\section{Introduction}

Thermodynamics is built around the concept of thermodynamic entropy,
while statistical mechanics provides its microscopic counterpart in
the form of Boltzmann entropy. For equilibrium systems, the relationships
among thermodynamic entropy, Boltzmann entropy, and the canonical
distribution are well established \cite{Boltzmann1896,Gibbs1902,Jaynes_1957,Jaynes_1957_II}.
In nonequilibrium systems, however, and especially in open systems,
nonequilibrium steady states, and stochastic dynamics, the meaning
of entropy becomes more subtle.

Several entropy-related quantities then have to be distinguished.
The local self-information $S_{\mathrm{Sh}}(x,t)=-\log\phi(x,t)$,
its ensemble average known as the Shannon entropy $\langle S_{\mathrm{Sh}}\rangle$
\cite{Shannon_1948a,Shannon_1948b}, an entropy potential $S_{\mathrm{P}}$
generating nonequilibrium gradient flows, the Onsager--Machlup--Hashitsume
action $S_{\mathrm{OMH}}$ \cite{Hashitsume_1952,Onsager_1953,Machlup_1953,Hashitsume_1956},
and the thermodynamic entropy $S_{\mathrm{th}}$ used in equilibrium
thermodynamics are closely related, but they are not identical.

\LyXZeroWidthSpace{}

In recent years, we have developed Nambu nonequilibrium thermodynamics
(NNET) as a framework for describing systems far from equilibrium
using Nambu brackets \cite{katagiri2022fluctuating,katagiri2025nambu1,katagiri2025nambu2,katagiri2025nambu3}.
In NNET, the time evolution is decomposed into a nondissipative, volume-preserving
sector generated by the Nambu bracket and a dissipative sector generated
by the gradient of an entropy-like potential. This decomposition makes
it possible to describe oscillatory far-from-equilibrium dynamics,
such as the Belousov-{}-Zhabotinsky reaction, as well as spiking
and chaotic behavior, in terms of the interplay between circulating
Nambu flows and dissipative relaxation. The Nambu bracket is a generalization
of the Poisson bracket. In analogy with Hamiltonian flow generated
by the Poisson bracket in Hamiltonian mechanics, the flow generated
by the Nambu bracket in Nambu mechanics is referred to as Nambu flow.

\LyXZeroWidthSpace{}

In particular, our analysis of an open gas--piston system coupled
to a thermal bath and an external-pressure reservoir demonstrated
that the thermodynamic entropy of the open subsystem, $S_{\mathrm{th}}$\LyXZeroWidthSpace ,
and the NNET dissipation potential, $S_{\mathrm{NB}}$, are physically
distinct quantities \cite{NNET-Piston}. The thermodynamic entropy
$S_{\mathrm{th}}$ may exhibit nonmonotonic oscillations as a result
of heat and work exchange with the thermal and pressure reservoirs.
By contrast, the bath-relative dissipation potential

\begin{equation}
{\normalsize S_{\mathrm{NB}}=S_{\mathrm{th}}-\frac{H_{1}}{T_{\mathrm{b}}}}
\end{equation}

becomes a Lyapunov function when it is invariant under the nondissipative
Nambu flow and the irreversible dynamics closes as a positive-semidefinite
gradient flow generated by $S_{\mathrm{NB}}$. Throughout this paper,
we use $S_{\mathrm{NB}}$ consistently to denote the dissipation potential,
without introducing additional notation. The expression above adopts
a convention in which $S_{\mathrm{NB}}$ has the dimensions of entropy.
These results show that $S_{\mathrm{NB}}$\LyXZeroWidthSpace{} should
not be identified with the thermodynamic entropy itself, but should
instead be understood as a Massieu- or free-energy-like potential
governing irreversible relaxation in an open system.

\LyXZeroWidthSpace{}

The purpose of this paper is to clarify how $S_{\mathrm{NB}}$ enters
the nonequilibrium stationary distribution of stochastic NNET. Specifically,
we show that, when the Nambu flow preserves $S_{\mathrm{NB}}$ and
an Einstein-type relation holds between the diffusion matrix and the
transport tensor \cite{Einstein_1905_Brownian}, the stationary probability
density takes the form

\begin{equation}
\phi_{\mathrm{st}}(x)=\frac{1}{Z_{\mathrm{NB}}}\exp\left[\frac{S_{\mathrm{NB}}}{k_{\mathrm{B}}}\right].
\end{equation}

The emergence of exponential stationary distributions of this type
has previously been discussed by Frank \cite{Frank_2016}. The central
point of the present work is more specific: the dissipation potential
$S_{\mathrm{NB}}$, derived independently from the thermodynamics
of an open system, can be identified as the stationary quasipotential
governing the nonequilibrium stationary distribution. Indeed, the
local self-information satisfies 
\begin{equation}
S_{\mathrm{Sh}}(x)=-\ln\phi_{\mathrm{st}}=\ln Z_{\mathrm{NB}}-\frac{S_{\mathrm{NB}}(x)}{k_{\mathrm{B}}}.
\end{equation}

Thus, $S_{\mathrm{NB}}$ is not the Shannon entropy itself. Rather,
it determines the logarithmic statistical weight of the stationary
distribution and thereby plays the role of a stationary quasipotential.

\LyXZeroWidthSpace{}

For clarity, the entropy-related quantities used throughout this paper
are summarized in Table \ref{tab:entropy-correspondence}.

\begin{table}[H]
\centering{}%
\begin{tabular}{|c|c|c|c|}
\hline 
{\tiny\textbf{\textsc{Symbol}}}  & {\tiny\textbf{\textsc{Name}}}  & {\tiny\textbf{\textsc{Role}}}  & {\tiny\textbf{\textsc{Remarks}}}\tabularnewline
\hline 
{\tiny\textbf{\textsc{$S_{\mathrm{th}}$}}}  & {\scriptsize\textbf{\textsc{Thermodynamic entropy}}}  & {\scriptsize\textbf{\textsc{Thermodynamic state function}}}  & {\scriptsize\textbf{\textsc{Generally nonmonotonic in open systems}}}\tabularnewline
\hline 
{\tiny\textbf{\textsc{$S_{\mathrm{Sh}}$}}}  & {\scriptsize\textbf{\textsc{Self-information}}}  & {\scriptsize\textbf{\textsc{Local quantity defined by $-\ln\phi$}}}  & {\scriptsize\textbf{\textsc{Its expectation value is the Shannon entropy}}}\tabularnewline
\hline 
{\tiny\textbf{\textsc{$S_{\mathrm{P}}$}}}  & {\scriptsize\textbf{\textsc{Dissipation potential}}}  & {\scriptsize\textbf{\textsc{Potential generating a gradient flow}}}  & {\tiny\textbf{\textsc{Not necessarily identical to the thermodynamic
entropy}}}\tabularnewline
\hline 
{\tiny\textbf{\textsc{$S_{\mathrm{NB}}$}}}  & {\scriptsize\textbf{\textsc{Nambu-type dissipation potential}}}  & {\tiny\textbf{\textsc{Generates the irreversible sector of NNET}}}  & {\scriptsize\textbf{\textsc{Main object of the present work}}}\tabularnewline
\hline 
{\tiny\textbf{\textsc{$S_{\mathrm{OMH}}$}}}  & {\tiny\textbf{\textsc{Onsager--Machlup--Hashitsume action}}}  & {\scriptsize\textbf{\textsc{Action governing path probabilities}}}  & {\tiny\textbf{\textsc{Related to the self-information as an effective
action}}}\tabularnewline
\hline 
\end{tabular}\caption{Correspondence among the entropy-related quantities used in this work.}\label{tab:entropy-correspondence}
\end{table}

The remainder of this paper is organized as follows. In Sec. II, we
clarify the relation between self-information and Shannon entropy
using the pure diffusion equation as the simplest example. In Sec.
III, we derive the stationary distribution for Langevin and Fokker--Planck
systems driven by a gradient potential. In Sec. IV, we discuss stationary
distributions and stationary probability currents in the presence
of Hamiltonian-type rotational flows. In Sec. V, we extend the same
analysis to NNET and derive the Nambu-type stationary distribution
together with the statistical interpretation of $S_{\mathrm{NB}}$\LyXZeroWidthSpace .
In Sec. VI, we reformulate $S_{\mathrm{NB}}$ for open systems in
the language of statistical ensembles and introduce both the saddle-point
approximation and generalized thermodynamic forces. In Sec. VII, we
discuss the physical meaning and range of validity of the results,
possible q-generalizations, and the limitations of the present formulation
for strongly nonlinear systems. Finally, Appendix A presents a preliminary
discussion of path-probability ratios and fluctuation theorems.

\section{Diffusion Equation and Self-Information}

\subsection{Pure diffusion}

As the simplest example of stochastic dynamics, we first consider
the stochastic process driven by Gaussian white noise,

\begin{equation}
\dot{x}(t)=\xi(t),
\end{equation}

where 
\begin{equation}
\langle\xi(t)\rangle=0,\qquad\langle\xi(t)\xi(t')\rangle=2D\delta(t-t')
\end{equation}

The corresponding probability density $\phi(x,t)$ obeys the diffusion
equation

\begin{equation}
\frac{\partial\phi}{\partial t}=D\frac{\partial^{2}\phi}{\partial x^{2}}
\end{equation}

In a stationary state,

\begin{equation}
\frac{\partial\phi}{\partial t}=0,
\end{equation}

and hence, in one dimension,

\begin{equation}
\phi(x)=Ax+B.
\end{equation}

For a closed system on a finite domain with periodic boundary conditions,
periodicity requires $A=0$. The stationary probability density is
therefore uniform.

\LyXZeroWidthSpace{}

By contrast, consider a finite interval $0\le x\le L$ with fixed
boundary values,

\begin{equation}
\phi(0)=\phi_{0},\qquad\phi(L)=\phi_{L}.
\end{equation}

The stationary solution is then

\begin{equation}
\phi(x)=Ax+B,\ A=\frac{\phi_{L}-\phi_{0}}{L},\ B=\phi_{0}.
\end{equation}

Thus, a nonuniform linear stationary profile is allowed. Introducing
the probability current through the continuity equation,

\begin{equation}
\frac{\partial\phi}{\partial t}+\frac{\partial J}{\partial x}=0,
\end{equation}

one finds

\begin{equation}
J=-D\frac{\partial\phi}{\partial x}=-D\frac{\phi_{L}-\phi_{0}}{L}.
\end{equation}

In general, $J\neq0$. Therefore, even though the probability density
is time independent,

\begin{equation}
\partial_{t}\phi=0,
\end{equation}

a finite stationary probability current may persist. This provides
the simplest example of a nonequilibrium steady state in which the
boundaries act as two reservoirs fixing different values of the probability
density.

\subsection{Hamilton--Jacobi-type equation for self-information}

We define the local self-information associated with the probability
density $\phi(x,t)$ by

\begin{equation}
S_{\mathrm{Sh}}(x,t)=-\ln\phi(x,t),\qquad\phi(x,t)=e^{-S_{\mathrm{Sh}}(x,t)}.
\end{equation}

Here, $S_{\mathrm{Sh}}(x,t)$ denotes the local self-information (or
surprisal) associated with the probability density $\phi(x,t)$, rather
than the Boltzmann entropy $S_{B}=k_{B}\ln\Omega$ defined in terms
of the density of microscopic states. Its ensemble average gives the
Shannon entropy, as discussed below.

Substituting $\phi=e^{-S_{\mathrm{Sh}}}$ into the diffusion equation,
we obtain 
\begin{equation}
\frac{\partial S_{\mathrm{Sh}}}{\partial t}=-D\left(\frac{\partial S_{\mathrm{Sh}}}{\partial x}\right)^{2}+D\frac{\partial^{2}S_{\mathrm{Sh}}}{\partial x^{2}}
\end{equation}

The first term on the right-hand side is a nonlinear Hamilton--Jacobi-type
term, whereas the second is a viscosity term originating from diffusion.
Thus, the evolution equation for $S_{\mathrm{Sh}}$ may be viewed
as a viscous Hamilton--Jacobi equation.

\LyXZeroWidthSpace{}

This equation also shows that the local self-information $S_{\mathrm{Sh}}(x,t)$
is not necessarily monotonic in time. At a given point $x$, it increases
when 
\begin{equation}
\left(\partial_{x}S_{\mathrm{Sh}}\right)^{2}<\partial_{x}^{2}S_{\mathrm{Sh}},
\end{equation}

whereas it decreases when

\begin{equation}
\left(\partial_{x}S_{\mathrm{Sh}}\right)^{2}>\partial_{x}^{2}S_{\mathrm{Sh}}.
\end{equation}

It is therefore essential to distinguish the local self-information
from its ensemble average, namely the Shannon entropy.

\subsection{Shannon entropy and relative entropy}

The Shannon entropy is defined as the ensemble average of the self-information,

\begin{equation}
\langle S_{\mathrm{Sh}}(t)\rangle=-\int\phi(x,t)\log\phi(x,t)\,dx=\int S_{\mathrm{Sh}}(x,t)\phi(x,t)\,dx
\end{equation}

For closed or periodic boundary conditions, or sufficiently rapid
decay such that the boundary terms vanish, the diffusion equation
gives 
\begin{equation}
\frac{d}{dt}\langle S_{\mathrm{Sh}}\rangle=D\int\phi\left(\partial_{x}\log\phi\right)^{2}dx=D\int\phi\left(\partial_{x}S_{\mathrm{Sh}}\right)^{2}dx\ge0.
\end{equation}

Thus, in a closed system governed by pure diffusion, the Shannon entropy
increases monotonically and is ultimately maximized by the uniform
distribution.

\LyXZeroWidthSpace{}

This conclusion, however, depends on the boundary conditions. When
the boundary values are fixed by concentration reservoirs, as in the
preceding subsection, the stationary distribution is nonuniform and
supports a nonzero probability current. In this case, the natural
quantity characterizing relaxation is not the Shannon entropy itself,
but the relative entropy, or Kullback--Leibler divergence, with respect
to the stationary distribution $\phi_{\mathrm{st}}:$ 
\begin{equation}
D_{\mathrm{KL}}(\phi||\phi_{\mathrm{st}})=\int\phi(x,t)\log\frac{\phi(x,t)}{\phi_{\mathrm{st}}(x)}\,dx.
\end{equation}

As we will see below, this relative entropy plays a central role as
a measure of relaxation toward nonequilibrium stationary distributions.

\subsection{Relation to the Onsager--Machlup--Hashitsume action}

The diffusion process introduced above can also be represented in
a path-integral form. With the noise correlation normalized as

\begin{equation}
\langle\xi(t)\xi(t')\rangle=2D\delta(t-t')
\end{equation}

the transition probability can be written formally as

\begin{equation}
\phi(x,t|x_{0},0)=\int_{x(0)=x_{0}}^{x(t)=x}\mathcal{D}x\,\exp\left[-\frac{1}{4D}\int_{0}^{t}\dot{x}^{2}\,dt'\right].
\end{equation}

Here and below, normalization factors and Jacobian corrections associated
with changes of variables are omitted. The functional appearing in
the exponential is the Onsager--Machlup--Hashitsume-type action
for this simple diffusion process,

\begin{equation}
S_{\mathrm{OMH}}[x]=\frac{1}{4D}\int_{0}^{t}\dot{x}^{2}dt'.
\end{equation}

Including the initial probability density $\phi(x_{0},0)$, the probability
density at time $t$ is

\begin{equation}
\phi(x,t)=\int dx_{0}\,\phi(x,t|x_{0},0)\phi(x_{0},0)
\end{equation}

Accordingly, the local self-information becomes

\begin{equation}
S_{\mathrm{Sh}}(x,t)=-\ln\left[\int dx_{0}\int_{x(0)=x_{0}}^{x(t)=x}\mathcal{D}x\,\exp\left[-S_{\mathrm{OMH}}[x]\right]\phi(x_{0},0)\right].
\end{equation}

In this sense, the self-information may be regarded as an effective
action obtained by integrating the OMH path action over all trajectories
compatible with the prescribed endpoint. In the stochastic NNET considered
below, the deterministic Nambu--gradient flow can likewise be interpreted
as the most probable path selected by the corresponding stochastic
path weight.

\section{Stationary Distribution for Gradient-Potential Systems}

\subsection{Langevin equation}

We next consider a Langevin equation driven by the gradient of a scalar
potential. In one dimension, the dynamics is given by

\begin{equation}
\dot{x}=L\frac{\partial S_{\mathrm{P}}}{\partial x}+\xi.
\end{equation}

Here, $S_{\mathrm{P}}$ is a scalar dissipation potential generating
the deterministic gradient-flow sector and need not coincide with
thermodynamic entropy. The coefficient $L>0$ is the transport coefficient,
which is assumed throughout this section to be independent of $x$
for simplicity. The noise is taken to be Gaussian and white, satisfying

\begin{equation}
\langle\xi(t)\rangle=0,\qquad\langle\xi(t)\xi(t')\rangle=2D\delta(t-t').
\end{equation}

\subsection{Fokker--Planck equation}

The Langevin dynamics introduced above is described by the corresponding
Fokker--Planck equation,

\begin{equation}
\frac{\partial\phi}{\partial t}=-\frac{\partial}{\partial x}\left(L\frac{\partial S_{\mathrm{P}}}{\partial x}\phi\right)+D\frac{\partial^{2}\phi}{\partial x^{2}},
\end{equation}

Defining the probability current as

\begin{equation}
J=L\frac{\partial S_{\mathrm{P}}}{\partial x}\phi-D\frac{\partial\phi}{\partial x},
\end{equation}

the Fokker--Planck equation takes the continuity-equation form 
\begin{equation}
\frac{\partial\phi}{\partial t}+\frac{\partial J}{\partial x}=0.
\end{equation}

\subsection{Exponential stationary distribution}

We first consider a stationary state satisfying detailed balance,
for which the probability current vanishes:

\begin{equation}
J=0.
\end{equation}

It then follows that 
\begin{equation}
L\frac{\partial S_{\mathrm{P}}}{\partial x}\phi_{\mathrm{st}}-D\frac{\partial\phi_{\mathrm{st}}}{\partial x}=0.
\end{equation}

Therefore, 
\begin{equation}
\partial_{x}\log\phi_{\mathrm{st}}=\frac{L}{D}\partial_{x}S_{\mathrm{P}},
\end{equation}

and the normalized stationary distribution is 
\begin{equation}
\phi_{\mathrm{st}}=\frac{1}{Z_{\mathrm{P}}}\exp\left[\frac{L}{D}S_{\mathrm{P}}\right],
\end{equation}

where $Z_{\mathrm{P}}$ is the normalization constant. The corresponding
self-information is

\begin{equation}
S_{\mathrm{Sh}}(x)=-\log\phi_{\mathrm{st}}(x)=\log Z_{\mathrm{P}}-\frac{L}{D}S_{\mathrm{P}}(x).
\end{equation}

We now impose the Einstein-type relation 
\begin{equation}
D=k_{\mathrm{B}}L.
\end{equation}

Under this relation, the stationary distribution takes the form

\begin{equation}
\phi_{\mathrm{st}}=\frac{1}{Z_{\mathrm{P}}}\exp\left[\frac{S_{\mathrm{P}}}{k_{\mathrm{B}}}\right].
\end{equation}

Accordingly,

\begin{equation}
S_{\mathrm{Sh}}(x)=\log Z_{\mathrm{P}}-\frac{S_{\mathrm{P}}(x)}{k_{\mathrm{B}}}.
\end{equation}

Taking the ensemble average gives 
\begin{equation}
\langle S_{\mathrm{Sh}}\rangle=\log Z_{\mathrm{P}}-\frac{\langle S_{\mathrm{P}}\rangle}{k_{\mathrm{B}}}.
\end{equation}

This relation connects the local self-information to the entropy potential
in the stationary state. It does not imply that $S_{\mathrm{Sh}}$
and $S_{\mathrm{P}}$ are identical, nor that the same relation holds
for an arbitrary time-dependent probability distribution.

\subsection{Monotonic decay of the Kullback--Leibler divergence}

Relaxation toward the stationary distribution can be characterized
by the monotonic decay of the relative entropy, or Kullback--Leibler
divergence, 
\begin{equation}
D_{\mathrm{KL}}(\phi||\phi_{\mathrm{st}})=\int\phi(x,t)\log\frac{\phi(x,t)}{\phi_{\mathrm{st}}(x)}\,dx
\end{equation}

When $D$ is constant and the boundary terms vanish, the Fokker--Planck
equation can be rewritten as

\begin{equation}
\frac{\partial\phi}{\partial t}=D\frac{\partial}{\partial x}\left[\phi\frac{\partial}{\partial x}\log\frac{\phi}{\phi_{\mathrm{st}}}\right].
\end{equation}

It then follows that 
\begin{equation}
\frac{d}{dt}D_{\mathrm{KL}}(\phi||\phi_{\mathrm{st}})=-D\int\phi\left(\partial_{x}\log\frac{\phi}{\phi_{\mathrm{st}}}\right)^{2}dx\le0.
\end{equation}

Thus, whenever a stationary distribution $\phi_{\mathrm{st}}$\LyXZeroWidthSpace{}
exists, the time-dependent distribution $\phi$ approaches it in the
sense of relative entropy.

\LyXZeroWidthSpace{}

A qualification is required regarding the potential $S_{\mathrm{P}}$
itself. When the stochastic differential equation is treated rigorously
in the Itô convention, the evolution of the ensemble average $\langle S_{\mathrm{P}}(x_{t})\rangle$
contains an Itô correction proportional to

\begin{equation}
D\partial_{x}^{2}S_{\mathrm{P}}.
\end{equation}

Consequently, $\langle S_{\mathrm{P}}\rangle$ is not guaranteed to
increase monotonically in the presence of fluctuations. The relevant
monotonic quantity in the present analysis is therefore not $S_{\mathrm{P}}$
itself, but the Kullback--Leibler divergence relative to the stationary
distribution, which decreases monotonically under the conditions stated
above.

\section{Stationary Distributions in the Presence of Rotational Flows}

\subsection{Hamiltonian current}

We next consider a system in which a nongradient rotational flow coexists
with the gradient flow. In two dimensions, let the stochastic dynamics
be

\begin{equation}
\dot{x}^{i}=N^{i}(x)+L^{ij}\partial_{j}S_{\mathrm{P}}+\xi^{i},\qquad N^{i}=\epsilon^{ij}\partial_{j}H.
\end{equation}

The Hamiltonian flow $N^{i}$ is volume preserving, since

\begin{equation}
\partial_{i}N^{i}=0.
\end{equation}

The noise is assumed to satisfy 
\begin{equation}
\langle\xi^{i}(t)\xi^{j}(t')\rangle=2D^{ij}\delta(t-t').
\end{equation}

The corresponding Fokker--Planck equation is

\begin{equation}
\frac{\partial\phi}{\partial t}=-\partial_{i}\left[(N^{i}+L^{ij}\partial_{j}S_{\mathrm{P}})\phi\right]+\partial_{i}\left(D^{ij}\partial_{j}\phi\right).
\end{equation}

The associated probability current is therefore

\begin{equation}
J^{i}=(N^{i}+L^{ij}\partial_{j}S_{\mathrm{P}})\phi-D^{ij}\partial_{j}\phi.
\end{equation}

\subsection{Condition $N\cdot\nabla S_{\mathrm{P}}=0$}

We assume that the diffusion and transport tensors satisfy the Einstein-type
relation

\begin{equation}
D^{ij}=k_{\mathrm{B}}L^{ij}.
\end{equation}

The stationary distribution generated by the gradient sector alone
is then

\begin{equation}
\phi_{\mathrm{st}}=\frac{1}{Z_{\mathrm{P}}}\exp\left[\frac{S_{\mathrm{P}}}{k_{\mathrm{B}}}\right].
\end{equation}

A sufficient condition for this distribution to remain stationary
in the presence of the rotational flow is

\begin{equation}
N^{i}\partial_{i}S_{\mathrm{P}}=0
\end{equation}

Indeed, under this condition, 
\begin{equation}
N^{i}\partial_{i}\phi_{\mathrm{st}}=\frac{\phi_{\mathrm{st}}}{k_{\mathrm{B}}}N^{i}\partial_{i}S_{\mathrm{P}}=0.
\end{equation}

Since the Hamiltonian flow is divergence free,

\begin{equation}
\partial_{i}N^{i}=0,
\end{equation}

it follows that

\begin{equation}
\partial_{i}(N^{i}\phi_{\mathrm{st}})=\partial_{i}(N^{i})\phi_{\mathrm{st}}+N^{i}\partial_{i}\phi_{\mathrm{st}}=0.
\end{equation}

Thus, the rotational current does not alter the stationary probability
density.

\LyXZeroWidthSpace{}

In particular, when the entropy potential depends on the Hamiltonian
only,

\begin{equation}
S_{\mathrm{P}}=S_{\mathrm{P}}(H),
\end{equation}

the condition is automatically satisfied. The Hamiltonian flow is
tangent to the level sets of $H$, so that

\begin{equation}
N\cdot\nabla H=0
\end{equation}

and therefore 
\begin{equation}
N\cdot\nabla S_{\mathrm{P}}=\frac{dS_{\mathrm{P}}}{dH}N\cdot\nabla H=0.
\end{equation}

\subsection{Reversible stationary probability current}

When the conditions above are satisfied, the gradient and diffusion
contributions to the probability current cancel each other in the
stationary state. The remaining stationary current is therefore

\begin{equation}
J_{\mathrm{st}}^{i}=N^{i}\phi_{\mathrm{st}}=\epsilon^{ij}\partial_{j}H\,\frac{1}{Z_{\mathrm{P}}}\exp\left[\frac{S_{\mathrm{P}}(H)}{k_{\mathrm{B}}}\right].
\end{equation}

This current satisfies 
\begin{equation}
\partial_{i}J_{\mathrm{st}}^{i}=0,
\end{equation}

but is generally nonzero: 
\begin{equation}
J_{\mathrm{st}}^{i}\ne0.
\end{equation}

The stationary state therefore does not satisfy detailed balance and
should be regarded as a nonequilibrium steady state carrying a circulating
probability current.

\LyXZeroWidthSpace{}

This stationary current is generated by the reversible Hamiltonian
sector and does not itself represent dissipative relaxation. Indeed,
in the time evolution of the relative entropy, the Hamiltonian contribution
vanishes because 
\begin{equation}
\partial_{i}N^{i}=0,\ N^{i}\partial_{i}S_{\mathrm{P}}=0,
\end{equation}

together with the assumed boundary conditions. Relaxation toward the
stationary distribution is consequently governed entirely by the gradient
and diffusion sectors. Thus, the rotational flow sustains a nonvanishing
probability current in the stationary state, while leaving the rate
of relative-entropy dissipation toward that state unchanged.

\section{Nambu-Type Stationary Distribution in NNET}

\subsection{Nambu--Langevin equation}

We consider a three-dimensional Langevin equation of the NNET form.
Let the state variables be

\begin{equation}
x=(x_{1},\ x_{2},\ x_{3}),
\end{equation}

and write the stochastic dynamics as 
\begin{equation}
\dot{x}^{i}=N^{i}(x)+L^{ij}(x)\partial_{j}S_{\mathrm{NB}}+\xi^{i},\qquad N^{i}=\epsilon^{ijk}\partial_{j}H_{1}\partial_{k}H_{2}.
\end{equation}

The first term is the Nambu flow generated by the two Hamiltonians
$H_{1}$ and $H_{2}$, whereas the second is the gradient flow generated
by the dissipation potential $S_{\mathrm{NB}}.$ We assume that the
noise satisfies 
\begin{equation}
\langle\xi^{i}(t)\xi^{j}(t')\rangle=2D^{ij}(x)\delta(t-t').
\end{equation}

For state-dependent diffusion, the noise may be represented by introducing
a noise-amplitude matrix, or vielbein, $e_{a}^{i}(x):$

\begin{equation}
\xi^{i}(t)=e_{a}^{i}(x)\eta^{a}(t),\ \langle\eta^{a}(t)\eta^{b}(t')\rangle=\delta^{ab}\delta(t-t'),
\end{equation}

with 
\begin{equation}
D^{ij}(x)=\frac{1}{2}\delta^{ab}e_{a}^{i}(x)e_{b}^{j}(x).
\end{equation}

It then follows that 
\begin{equation}
\langle\xi^{i}(t)\xi^{j}(t')\rangle=\delta^{ab}e_{a}^{i}e_{b}^{j}\delta(t-t')=2D^{ij}\delta(t-t').
\end{equation}

Following the Onsager--Machlup--Hashitsume path-integral formulation
\cite{Hashitsume_1952,Onsager_1953,Machlup_1953,Hashitsume_1956},
the corresponding Fokker--Planck equation for stochastic NNET is
written as

\begin{equation}
\frac{\partial\phi}{\partial t}=-\partial_{i}\left[(N^{i}+L^{ij}(x)\partial_{j}S_{\mathrm{NB}})\phi\right]+\partial_{i}\left[D^{ij}(x)\partial_{j}\phi\right].
\end{equation}

Accordingly, the probability current is 
\begin{equation}
J^{i}=(N^{i}+L^{ij}\partial_{j}S_{\mathrm{NB}})\phi-D^{ij}\partial_{j}\phi.
\end{equation}

For multiplicative noise, the stochastic convention is chosen such
that the diffusion term takes the divergence form 
\begin{equation}
\partial_{i}\left(D^{ij}\partial_{j}\phi\right).
\end{equation}

\subsection{Stationary distribution in stochastic Nambu nonequilibrium thermodynamics}

We now derive the stationary distribution of stochastic Nambu nonequilibrium
thermodynamics. We assume that the probability density is defined
with respect to the standard volume measure

\begin{equation}
d^{3}x=dx^{1}dx^{2}dx^{3}.
\end{equation}

Let the transport and diffusion tensors satisfy the Einstein-type
relation 
\begin{equation}
D^{ij}(x)=k_{\mathrm{B}}L^{ij}(x).
\end{equation}

We further impose boundary conditions for which the relevant boundary
terms vanish, such as periodic boundary conditions, sufficiently rapid
decay at infinity, or no-flux boundary conditions.

\LyXZeroWidthSpace{}

By the antisymmetry of the Nambu bracket, the Nambu flow satisfies

\begin{equation}
\partial_{i}N^{i}=0,\qquad N^{i}\partial_{i}H_{1}=N^{i}\partial_{i}H_{2}=0.
\end{equation}

Consequently, when the dissipation potential can be expressed as 
\begin{equation}
S_{\mathrm{NB}}=S_{\mathrm{NB}}(H_{1},H_{2}),
\end{equation}

it is preserved by the Nambu flow: 
\begin{equation}
N^{i}\partial_{i}S_{\mathrm{NB}}=0.
\end{equation}

Under this condition, the normalized distribution 
\begin{equation}
\phi_{\mathrm{st}}=\frac{1}{Z_{\mathrm{NB}}}\exp\left[\frac{S_{\mathrm{NB}}(H_{1},H_{2})}{k_{\mathrm{B}}}\right]
\end{equation}

is stationary, provided that the normalization factor $Z_{\mathrm{NB}}$
is finite. Indeed, 
\begin{equation}
\text{\ensuremath{\partial}}_{j}\phi_{\mathrm{st}}=\frac{1}{k_{\mathrm{B}}}\phi_{\mathrm{st}}\partial_{j}S_{\mathrm{NB}}.
\end{equation}

More generally, the representation $S_{\mathrm{NB}}=S_{\mathrm{NB}}(H_{1},H_{2})$
is a sufficient condition; the essential requirement is

\begin{equation}
N\cdot\nabla S_{\mathrm{NB}}=0.
\end{equation}

Using the Einstein-type relation, we then obtain

\begin{equation}
L^{ij}\partial_{j}S_{\mathrm{NB}}\phi_{\mathrm{st}}-D^{ij}\partial_{j}\phi_{\mathrm{st}}=0.
\end{equation}

Thus, the dissipative drift and diffusion currents cancel in the stationary
state, leaving the stationary probability current

\begin{equation}
J_{\mathrm{st}}^{i}=N^{i}\phi_{\mathrm{st}}=\epsilon^{ijk}\partial_{j}H_{1}\partial_{k}H_{2}\,\frac{1}{Z_{\mathrm{NB}}}\exp\left(\frac{S_{\mathrm{NB}}(H_{1},H_{2})}{k_{\mathrm{B}}}\right)
\end{equation}

Its divergence vanishes: 
\begin{equation}
\begin{aligned}\partial_{i}J_{\mathrm{st}}^{i} & =\left(\partial_{i}N^{i}\right)\phi_{\mathrm{st}}+N^{i}\partial_{i}\phi_{\mathrm{st}}\\
 & =\frac{1}{k_{\mathrm{B}}}\phi_{\mathrm{st}}N^{i}\partial_{i}S_{\mathrm{NB}}=0.
\end{aligned}
\end{equation}

The probability density is therefore stationary even though a circulating
probability current remains along the Nambu flow. Fokker--Planck
equations supporting this type of stationary conservative current
have also been discussed as nonequilibrium diffusion equations with
an additional divergence-free flow \cite{Frank_2016}.

\subsection{Relation between $S_{\mathrm{Sh}}$ and $S_{\mathrm{NB}}$}

The stationary distribution immediately gives the local self-information
as

\begin{equation}
S_{\mathrm{Sh}}(x)=-\log\phi_{\mathrm{st}}(x)=\log Z_{\mathrm{NB}}-\frac{S_{\mathrm{NB}}(x)}{k_{\mathrm{B}}}.
\end{equation}

Taking the ensemble average yields

\begin{equation}
\langle S_{\mathrm{Sh}}\rangle=\log Z_{\mathrm{NB}}-\frac{\langle S_{\mathrm{NB}}\rangle}{k_{\mathrm{B}}}.
\end{equation}

Thus, in the nonequilibrium stationary state, $S_{\mathrm{NB}}$ is
related to the local self-information with the opposite sign, up to
the additive constant $\ln Z_{\mathrm{NB}}$. This relation does not
imply that $S_{\mathrm{NB}}$ is the Shannon entropy itself. Rather,
$S_{\mathrm{NB}}$ determines the logarithmic weight of the stationary
distribution and therefore acts as a local stationary quasipotential
for the self-information.

\subsection{Nonequilibrium stationary currents and relative entropy}

As in the preceding section, when 
\begin{equation}
N\cdot\nabla S_{\mathrm{NB}}=0,\ \partial_{i}N^{i}=0,
\end{equation}

the Nambu probability current preserves the stationary distribution.
Consequently, the Nambu contribution to the time evolution of the
relative entropy 
\begin{equation}
D_{\mathrm{KL}}(\phi||\phi_{\mathrm{st}})=\int\phi(x,t)\log\frac{\phi(x,t)}{\phi_{\mathrm{st}}(x)}\,dx
\end{equation}

vanishes up to boundary terms. When $D^{ij}(x)$ is symmetric and
positive semidefinite and the boundary terms vanish, the gradient
and diffusion sectors yield 
\begin{equation}
\frac{d}{dt}D_{\mathrm{KL}}(\phi||\phi_{\mathrm{st}})=-\int\phi\,\partial_{i}\log\frac{\phi}{\phi_{\mathrm{st}}}D^{ij}(x)\partial_{j}\log\frac{\phi}{\phi_{\mathrm{st}}}\,dx\le0.
\end{equation}

Thus, the Nambu flow generates a stationary circulating probability
current, whereas relaxation toward the stationary distribution is
governed by the dissipative drift and diffusion sectors.

\LyXZeroWidthSpace{}

The condition

\begin{equation}
N\cdot\nabla S_{\mathrm{NB}}=0
\end{equation}

is the stochastic counterpart of the deterministic NNET condition
\begin{equation}
\partial_{t}^{(H)}S_{\mathrm{NB}}=0.
\end{equation}

In other words, when the nondissipative Nambu flow preserves $S_{\mathrm{NB}}$
and the irreversible sector closes as a positive-semidefinite gradient
flow generated by $S_{\mathrm{NB}}$, the dissipation potential becomes
a Lyapunov function under these conditions. This structure is realized
explicitly in the open gas-piston system \cite{NNET-Piston}.

\section{$S_{\mathrm{NB}}$ in Open Systems and Statistical Ensembles}

\subsection{Roles of $S_{\mathrm{th}}$ and $S_{\mathrm{NB}}$ in the gas--piston
system}

We consider an open gas--piston system in which a gas exchanges heat
with a thermal bath at temperature $T_{\mathrm{b}}$, while a movable
piston is coupled to an external-pressure reservoir.

In the open gas--piston system, the thermodynamic entropy of the
subsystem, $S_{\mathrm{th}},$ is clearly distinguished from the dissipation
potential $S_{\mathrm{NB}}$ that generates irreversible relaxation
\cite{NNET-Piston}. Because the subsystem exchanges heat with a thermal
reservoir and work with a pressure reservoir, $S_{\mathrm{th}}$ may
exhibit nonmonotonic oscillations. The thermodynamic entropy of an
open subsystem is therefore not, in general, a Lyapunov function.

\LyXZeroWidthSpace{}

By contrast, in the gas-piston system, the bath-relative quantity

\begin{equation}
S_{\mathrm{NB}}=S_{\mathrm{th}}-\frac{H_{1}}{T_{\mathrm{b}}}
\end{equation}

acts as the dissipation potential generating the irreversible sector.
Here, $H_{1}$ is the extended energy, including the work term associated
with the pressure reservoir, and $T_{\mathrm{b}}$ is the temperature
of the thermal reservoir. Under this convention, $S_{\mathrm{NB}}$
has the dimensions of entropy.

\LyXZeroWidthSpace{}

When the nondissipative Nambu flow preserves the dissipation potential,

\begin{equation}
\partial_{t}^{(H)}S_{\mathrm{NB}}=0,
\end{equation}

and the irreversible sector can be written as

\begin{equation}
\partial_{t}^{(S)}x^{i}=L^{ij}\partial_{j}S_{\mathrm{NB}},\ L\succeq0,
\end{equation}

the time evolution of $S_{\mathrm{NB}}$ becomes 
\begin{equation}
\dot{S}_{NB}=L^{ij}\partial_{i}S_{\mathrm{NB}}\partial_{j}S_{\mathrm{NB}}\ge0.
\end{equation}

Crucially, this monotonicity does not follow from identifying $S_{\mathrm{NB}}$
with the thermodynamic entropy itself. Rather, it follows from two
structural conditions: the Nambu flow preserves $S_{\mathrm{NB}}$,
and the irreversible dynamics closes as a positive-semidefinite gradient
flow generated by $S_{\mathrm{NB}}$. In strongly nonlinear open systems,
such as the BZ reaction, these conditions need not hold globally.
In such cases, even the dissipation potential may exhibit nonmonotonic
oscillations.

\subsection{Massieu interpretation of $S_{\mathrm{NB}}=S_{\mathrm{th}}-H_{1}/T_{\mathrm{b}}$}

The dissipation potential introduced above naturally takes the form
of a Massieu-type potential when the system is coupled to a thermal
reservoir at fixed temperature $T_{\mathrm{b}}$. Indeed, 
\begin{equation}
S_{\mathrm{NB}}=S_{\mathrm{th}}-\frac{H_{1}}{T_{\mathrm{b}}}=-\frac{1}{T_{\mathrm{b}}}\left(H_{1}-T_{\mathrm{b}}S_{\mathrm{th}}\right).
\end{equation}

Thus, $S_{\mathrm{NB}}$ is the negative of the extended free energy
$H_{1}-T_{\mathrm{b}}S_{\mathrm{th}}$, divided by the reservoir temperature.
Maximizing the entropy-dimensional quantity $S_{\mathrm{NB}}$ is
therefore equivalent to minimizing the corresponding free energy under
fixed reservoir conditions. At thermal and mechanical equilibrium,
where $T=T_{\mathrm{b}}$ and $P=P_{\mathrm{b}}$, this quantity reduces
to the ordinary Massieu-{}-Planck potential of the corresponding
equilibrium ensemble. Thus, the terminology ``Massieu-type'' away
from equilibrium is consistent with the standard equilibrium thermodynamic
structure.

\LyXZeroWidthSpace{}

For an open system coupled additionally to chemical reservoirs, an
analogous expression is

\begin{equation}
S_{\mathrm{NB}}=S_{\mathrm{th}}-\frac{1}{T_{\mathrm{b}}}\left(U-\sum_{a}\mu_{a}^{(\mathrm{b})}N_{a}\right),
\end{equation}

where $\mu_{a}^{(\mathrm{b})}$ denotes the chemical potential of
species $a$ imposed by the external reservoir. This quantity is a
grand-potential-type Massieu function appropriate to a system coupled
to both thermal and chemical reservoirs, and therefore provides a
natural candidate for the dissipation potential of an open chemical
reaction system \cite{deGroot1962,GlansdorffPrigogine1971,KondepudiPrigogine2014}.

\LyXZeroWidthSpace{}

In a general NNET system, however, $S_{\mathrm{NB}}$ need not always
possess such a Massieu form. When it is constructed from a Helmholtz
decomposition and a local Darboux representation, $S_{\mathrm{NB}}$
is defined primarily as the scalar potential generating the gradient
component of a given velocity field. It coincides with a free-energy-like
potential constructed from thermodynamic state variables and reservoir
parameters only when the reservoir structure is well defined and the
dissipative sector closes as the gradient flow of that potential.

\subsection{Nambu canonical distribution}

\subsubsection{Review of the Boltzmann--canonical correspondence at equilibrium}

We first recall the corresponding structure in equilibrium statistical
mechanics. Let $\Omega(E)$ denote the density of states at energy
$E$, and define the dimensionless Boltzmann entropy by 
\begin{equation}
s_{\mathrm{B}}(E)=\ln\Omega(E).
\end{equation}

The corresponding physical entropy is

\begin{equation}
S_{\mathrm{B}}(E)=k_{\mathrm{B}}s_{\mathrm{B}}(E).
\end{equation}

The canonical probability density for the energy is 
\begin{equation}
f(E;\beta)=\Omega(E)\frac{e^{-\beta E}}{Z(\beta)},
\end{equation}

and hence 
\begin{equation}
\ln f(E;\beta)=s_{\mathrm{B}}(E)-\beta E-\ln Z(\beta).
\end{equation}

Let 
\begin{equation}
U=\langle E\rangle
\end{equation}

be the mean energy, and write the thermodynamic free energy as 
\begin{equation}
F=U-TS_{\mathrm{th}}(U)=U-k_{\mathrm{B}}Ts(U)
\end{equation}

where $s(U)=S_{\mathrm{th}}(U)/k_{\mathrm{B}}$. Using 
\begin{equation}
\ln Z(\beta)=-\beta F=s(U)-\beta U,
\end{equation}

we obtain 
\begin{equation}
\ln f(E;\beta)=s_{\mathrm{B}}(E)-s(U)-\beta(E-U).
\end{equation}

We now expand $s_{\mathrm{B}}(E)$ about the saddle-point energy $E_{*}$,
defined as the stationary point of the canonical exponential weight.

\begin{equation}
\left.\frac{\partial s_{\mathrm{B}}}{\partial E}\right|_{E=E^{*}}=\beta,
\end{equation}

Then, 
\begin{equation}
\ln f(E;\beta)=\ln f(E_{*};\beta)+\frac{1}{2}\left.\frac{\partial^{2}s_{\mathrm{B}}}{\partial E^{2}}\right|_{E=E^{*}}(E-E^{*})^{2}+O((E-E_{*})^{3}).\label{eq:lnf(E)}
\end{equation}

In the thermodynamic limit, ensemble equivalence implies 
\begin{equation}
E_{*}\simeq U=\langle E\rangle.
\end{equation}

Identifying $E_{*}$ with $U$ to leading order in the thermodynamic
limit, Eq.\ref{eq:lnf(E)} becomes

\begin{equation}
\ln f(E;\beta)=\ln f(U;\beta)+\frac{1}{2}\left.\frac{\partial^{2}s_{\mathrm{B}}}{\partial E^{2}}\right|_{E=U}(E-U)^{2}+\cdots.
\end{equation}

Thermodynamic stability requires 
\begin{equation}
\partial_{E}^{2}s_{\mathrm{B}}|_{U}<0.
\end{equation}

Introducing the variance scale $\sigma^{2}$ through 
\begin{equation}
\left.\frac{\partial^{2}s_{\mathrm{B}}}{\partial E^{2}}\right|_{E=U}=-\frac{1}{\sigma^{2}},
\end{equation}

we find 
\begin{equation}
f(E;\beta)\simeq f(U;\beta)\exp\left[-\frac{(E-U)^{2}}{2\sigma^{2}}\right].
\end{equation}

Thus, the saddle-point approximation to the Massieu-type exponential
weight of the canonical ensemble yields Gaussian energy fluctuations
around equilibrium.

\subsubsection{Nambu canonical distribution}

We now carry this structure over to NNET. Introducing the dimensionless
potential

\begin{equation}
s_{\mathrm{NB}}(x)\equiv\frac{S_{\mathrm{NB}}(x)}{k_{\mathrm{B}}},
\end{equation}

we define the Nambu canonical distribution by 
\begin{equation}
p_{\mathrm{NB}}(x)=\frac{1}{Z_{\mathrm{NB}}}\exp s_{\mathrm{NB}}(x),\qquad Z_{\mathrm{NB}}=\int d\mu(x)\,\exp s_{\mathrm{NB}}(x),
\end{equation}

where $d\mu(x)$ denotes the measure on the state space. The corresponding
Massieu function is 
\begin{equation}
\Phi_{\mathrm{NB}}=\ln Z_{\mathrm{NB}},
\end{equation}

and the local self-information is 
\begin{equation}
S_{\mathrm{Sh}}(x)=-\log p_{\mathrm{NB}}(x)=\Phi_{\mathrm{NB}}-s_{\mathrm{NB}}(x).
\end{equation}

For the piston system, the dissipation potential is written as 
\begin{equation}
S_{\mathrm{NB}}=S_{\mathrm{th}}-\frac{H_{1}}{T_{\mathrm{b}}},
\end{equation}

and therefore has the dimensions of entropy. The corresponding dimensionless
potential is then 
\begin{equation}
s_{\mathrm{NB}}=\frac{S_{\mathrm{NB}}}{k_{\mathrm{B}}}.
\end{equation}

\subsection{Local Gaussian approximation near a nondegenerate stationary maximum}

Let $x^{*}$ denote a point at which $S_{\mathrm{NB}}$, or equivalently
$s_{\mathrm{NB}}$, attains its maximum. In the following, we assume
that $x^{*}$ is an isolated, nondegenerate maximum and that the Hessian
$\nabla^{2}s_{\mathrm{NB}}(x^{*})$ is negative definite. We further
assume that the normalization integral is dominated by a neighborhood
of $x^{*}$. Expanding $s_{\mathrm{NB}}$ about $x^{*}$, we obtain

\begin{equation}
s_{\mathrm{NB}}(x)=s_{\mathrm{NB}}(x^{*})+\frac{1}{2}\delta x^{T}\left[\nabla^{2}s_{\mathrm{NB}}(x^{*})\right]\delta x+O(\left\Vert \delta x\right\Vert ^{3}),
\end{equation}

where

\begin{equation}
\delta x=x-x^{*},\ \nabla s_{\mathrm{NB}}(x^{*})=0.
\end{equation}

Because the Nambu flow preserves $S_{\mathrm{NB}}$ if $S_{\mathrm{NB}}=S_{\mathrm{NB}}(H_{1},H_{2})$,
any Nambu orbit through a maximum remains on the same level set. Hence,
if the maximum is isolated, it must necessarily be a fixed point of
the Nambu flow.

The Nambu canonical distribution can therefore be approximated near
$x^{*}$ by

\begin{equation}
p_{\mathrm{NB}}(x)\simeq p_{\mathrm{NB}}(x^{*})\exp\left[-\frac{1}{2}\delta x^{T}C^{-1}\delta x\right],
\end{equation}

where the inverse covariance matrix is 
\begin{equation}
C^{-1}=-\nabla^{2}s_{\mathrm{NB}}(x^{*}).
\end{equation}

In terms of the dimensional dissipation potential $S_{\mathrm{NB}},$
the covariance matrix is 
\begin{equation}
C=-k_{\mathrm{B}}\left[\nabla^{2}S_{\mathrm{NB}}(x^{*})\right]^{-1}.
\end{equation}

Thus, within the saddle-point approximation, the NNET dissipation
potential determines the Gaussian fluctuation distribution around
the stationary state of the open system.

\subsubsection{Example: Gaussian fluctuations in the open gas--piston system}

As a concrete example, we consider Gaussian fluctuations near equilibrium
in an open gas--piston system. A gas of volume $V$ is confined by
a movable piston, and the system is coupled to a thermal reservoir
at temperature $T_{\mathrm{b}}$ and a pressure reservoir that imposes
a constant external pressure $P_{\mathrm{b}}$.

\LyXZeroWidthSpace{}

The deterministic NNET formulation of this system was discussed in
detail in Ref.\cite{NNET-Piston}. We take the state variables to
be 
\begin{equation}
\bm{x}=(V,\ p,\ S_{\mathrm{th}})
\end{equation}

where $p$ is the momentum of the piston and $S_{\mathrm{th}}$ is
the thermodynamic entropy of the gas enclosed by the piston. To obtain
explicit expressions, we assume below that the gas is ideal.

\LyXZeroWidthSpace{}

Following Ref. \cite{NNET-Piston}, we take the first Nambu Hamiltonian
$H_{1}$ to be the extended energy,

\begin{equation}
H_{1}=\frac{p^{2}}{2m}+U(S_{\mathrm{th}},V)+P_{\mathrm{b}}V
\end{equation}

and choose the thermodynamic entropy of the gas as the second Nambu
Hamiltonian,

\begin{equation}
H_{2}=S_{\mathrm{th}}
\end{equation}

These quantities are conserved in the nondissipative sector obtained
by excluding the irreversible gradient flow generated by $S_{\mathrm{NB}}$.

We introduce the following Massieu-type potential as the dissipation
potential $S_{\mathrm{NB}}$: 
\begin{equation}
S_{\mathrm{NB}}=S_{\mathrm{th}}-\frac{H_{1}}{T_{\mathrm{b}}}.
\end{equation}

Let the equilibrium point be 
\begin{equation}
x_{*}=(V_{*},\ 0,\ S_{*}).
\end{equation}

At this point, the equilibrium conditions

\begin{equation}
P(S_{*},V_{*})=P_{\mathrm{b}},\ T(S_{*},V_{*})=T_{\mathrm{b}}
\end{equation}

are satisfied, with $p_{*}=0$. Under these conditions, one can verify
that 
\begin{equation}
\nabla S_{\mathrm{NB}}(\bm{x}_{*})=0.
\end{equation}

Indeed, 
\begin{equation}
\frac{\partial S_{\mathrm{NB}}}{\partial V}=\frac{P-P_{\mathrm{b}}}{T_{\mathrm{b}}},
\end{equation}

\begin{equation}
\frac{\partial S_{\mathrm{NB}}}{\partial p}=-\frac{p}{mT_{\mathrm{b}}},
\end{equation}

and 
\begin{equation}
\frac{\partial S_{\mathrm{NB}}}{\partial S_{\mathrm{th}}}=1-\frac{T}{T_{\mathrm{b}}}.
\end{equation}

Substituting the equilibrium conditions shows that all three components
vanish.

\LyXZeroWidthSpace{}

For the ideal gas, we assume that the number of particles is fixed
and that the heat capacity at constant volume, $C_{\mathrm{V}}$,
is constant. Relative to the equilibrium point, the thermodynamic
entropy can then be written as

\begin{equation}
S_{\mathrm{th}}-S_{*}=C_{\mathrm{V}}\ln\frac{T}{T_{\mathrm{b}}}+R\ln\frac{V}{V_{*}},
\end{equation}

The internal energy and equation of state are given by

\begin{equation}
U=C_{\mathrm{V}}T,\ PV=RT.
\end{equation}

Here, $R$ denotes the gas constant appropriate to the amount of gas
contained in the system.

\LyXZeroWidthSpace{}

We define the deviation from equilibrium by

\begin{equation}
\delta\bm{x}=\bm{x}-\bm{x}_{*}=\left(\begin{array}{c}
\delta V\\
\delta p\\
\delta S_{\mathrm{th}}
\end{array}\right)
\end{equation}

Expanding $S_{\mathrm{NB}}$ about the equilibrium point to second
order gives

\begin{equation}
S_{\mathrm{NB}}(\bm{x})=S_{\mathrm{NB}}\left(\bm{x}_{*}\right)+\frac{1}{2}\delta\bm{x}^{\top}\nabla^{2}\left.S_{\mathrm{NB}}\right|_{*}\delta\bm{x}+O\left(\left\Vert \delta\bm{x}\right\Vert ^{3}\right).
\end{equation}

The linear term vanishes because 
\begin{equation}
\nabla S_{\mathrm{NB}}(\bm{x}_{*})=0.
\end{equation}

From the thermodynamic relations for an ideal gas, we obtain\footnote{From the entropy relation

\[
T(S_{\mathrm{th}},V)=T_{\mathrm{b}}\exp\!\left(\frac{S_{\mathrm{th}}-S_{*}}{C_{\mathrm{V}}}\right)\left(\frac{V}{V_{*}}\right)^{-R/C_{\mathrm{V}}},
\]

we obtain

\[
\left.\frac{\partial T}{\partial S_{\mathrm{th}}}\right|_{V,*}=\frac{T_{\mathrm{b}}}{C_{\mathrm{V}}}.
\]

Then, using $P=RT/V$ and differentiating with respect to $S_{\mathrm{th}}$
at fixed $V$, we obtain

\[
\left.\frac{\partial P}{\partial S_{\mathrm{th}}}\right|_{V,*}.
\]

Similarly,

\[
\left(\frac{\partial P}{\partial V}\right)_{S_{\mathrm{th}}}=\frac{\partial}{\partial V}\left(\frac{RT}{V}\right)
\]

gives

\[
\left.\frac{\partial P}{\partial V}\right|_{S_{\mathrm{th}},*}.
\]
} 
\begin{equation}
\left.\frac{\partial P}{\partial V}\right|_{S_{\mathrm{th}},*}=-\frac{RT_{\mathrm{b}}}{V_{*}^{2}}\left(1+\frac{R}{C_{\mathrm{V}}}\right),
\end{equation}

\begin{equation}
\left.\frac{\partial P}{\partial S_{\mathrm{th}}}\right|_{V,*}=\frac{RT_{\mathrm{b}}}{C_{\mathrm{V}}V_{*}},
\end{equation}

and

\begin{equation}
\left.\frac{\partial T}{\partial S_{\mathrm{th}}}\right|_{V,*}=\frac{T_{\mathrm{b}}}{C_{\mathrm{V}}}.
\end{equation}

Therefore, the Hessian of $S_{\mathrm{NB}}$ at equilibrium is

\begin{equation}
\left.\nabla^{2}S_{\mathrm{NB}}\right|_{*}=-\left(\begin{array}{ccc}
\frac{R}{V_{*}^{2}}\left(1+\frac{R}{C_{\mathrm{V}}}\right) & 0 & -\frac{R}{C_{\mathrm{V}}V_{*}}\\
0 & \frac{1}{mT_{\mathrm{b}}} & 0\\
-\frac{R}{C_{\mathrm{V}}V_{*}} & 0 & \frac{1}{C_{\mathrm{V}}}
\end{array}\right).
\end{equation}

Since $C_{\mathrm{V}}>0,R>0$ and $T_{\mathrm{b}}>0$, this Hessian
is negative definite. Hence, the equilibrium point $\bm{x}^{*}$ is
a nondegenerate local maximum of $S_{\mathrm{NB}}$.

\LyXZeroWidthSpace{}

The Nambu canonical distribution can therefore be approximated near
equilibrium by the Gaussian distribution 
\begin{equation}
p_{\mathrm{NB}}\left(\delta\bm{x}\right)\simeq\frac{1}{Z_{\mathrm{G}}}\exp\left[-\frac{1}{2}\delta\bm{x}^{\top}C^{-1}\delta\bm{x}\right].
\end{equation}

From the general result derived in Sec. 6.4, the covariance matrix
is given by

\begin{equation}
C=-k_{\mathrm{B}}\left(\left.\nabla^{2}S_{\mathrm{NB}}\right|_{*}\right)^{-1}.
\end{equation}

Inverting the Hessian above yields 
\begin{equation}
C=k_{\mathrm{B}}\left(\begin{array}{ccc}
\frac{V_{*}^{2}}{R} & 0 & V_{*}\\
0 & mT_{b} & 0\\
V_{*} & 0 & C_{\mathrm{V}}+R
\end{array}\right).
\end{equation}

Thus, the near-equilibrium fluctuations satisfy 
\begin{equation}
\langle\left(\delta V\right)^{2}\rangle=\frac{k_{\mathrm{B}}V_{*}^{2}}{R},
\end{equation}

\begin{equation}
\langle\left(\delta p\right)^{2}\rangle=mk_{\mathrm{B}}T_{\mathrm{b}},
\end{equation}

\begin{equation}
\langle\left(\delta S_{\mathrm{th}}\right)^{2}\rangle=k_{\mathrm{B}}\left(C_{\mathrm{V}}+R\right),
\end{equation}

and 
\begin{equation}
\langle\delta V\delta S_{\mathrm{th}}\rangle=k_{\mathrm{B}}V_{*}.
\end{equation}

By contrast, 
\begin{equation}
\langle\delta V\delta p\rangle=\langle\delta p\delta S_{\mathrm{th}}\rangle=0.
\end{equation}

The momentum fluctuation, 
\begin{equation}
\langle\left(\delta p\right)^{2}\rangle=mk_{\mathrm{B}}T_{\mathrm{b}}
\end{equation}

agrees with the usual equipartition result for the kinetic energy
of the piston. Moreover, the volume and thermodynamic-entropy fluctuations
exhibit a nonvanishing correlation, 
\begin{equation}
\langle\delta V\delta S_{\mathrm{th}}\rangle\neq0.
\end{equation}

This correlation is not introduced artificially; rather, it arises
from the thermodynamic coupling among volume, temperature, and entropy
encoded in the ideal-gas equation of state.

\LyXZeroWidthSpace{}

This example demonstrates that, in the open gas--piston system, the
same dissipation potential $S_{\mathrm{NB}}$ that generates deterministic
irreversible relaxation also determines the covariance of stationary
fluctuations near equilibrium. In other words, the local curvature
of $S_{\mathrm{NB}}$ directly determines both the magnitude and the
correlation structure of Gaussian fluctuations in the open system.
This provides a concrete realization of the general result of this
work, according to which $S_{\mathrm{NB}}$ acts as the quasipotential
governing the stationary distribution.

\subsubsection{van der Waals gas and approach to criticality}

We next consider the case of a van der Waals gas in the gas-{}-piston
system.

For a van der Waals gas, the thermodynamic entropy is given by

\begin{equation}
S_{\mathrm{th}}-S_{*}=C_{\mathrm{V}}\ln\frac{T}{T_{\mathrm{b}}}+R\ln\frac{V-b}{V^{*}-b},
\end{equation}

the internal energy by

\begin{equation}
U=C_{\mathrm{V}}T-\frac{a}{V},
\end{equation}

and the equation of state by

\begin{equation}
P=\frac{RT}{V-b}-\frac{a}{V^{2}}.
\end{equation}

Since the form of $S_{\mathrm{NB}}$ remains unchanged, its first
derivatives are independent of the specific equation of state. Therefore,
for the van der Waals gas, the equilibrium conditions

\begin{equation}
P=P_{\mathrm{b}},\ T=T_{\mathrm{b}},\ p=0
\end{equation}

again imply 
\begin{equation}
\nabla S_{\mathrm{NB}}=0.
\end{equation}

From the entropy relation, we obtain 
\begin{equation}
T\left(S_{\mathrm{th}},V\right)=T_{\mathrm{b}}\exp\left(\frac{S_{\mathrm{th}}-S_{*}}{C_{\mathrm{V}}}\right)\left(\frac{V-b}{V_{*}-b}\right)^{-R/C_{\mathrm{V}}},
\end{equation}

and hence

\begin{equation}
\left.\frac{\partial T}{\partial S_{\mathrm{th}}}\right|_{V,*}=\frac{T_{\mathrm{b}}}{C_{\mathrm{V}}}.
\end{equation}

Using the equation of state, we find

\begin{equation}
\left(\frac{\partial P}{\partial S_{\mathrm{th}}}\right)_{V}=\left(\frac{\partial}{\partial S_{\mathrm{th}}}\left(\frac{RT}{V-b}-\frac{a}{V^{2}}\right)\right)_{V}=\frac{RT}{C_{\mathrm{V}}\left(V-b\right)}.
\end{equation}
so that 
\begin{equation}
\left.\frac{\partial P}{\partial S_{\mathrm{th}}}\right|_{V,*}=\frac{RT_{\mathrm{b}}}{C_{\mathrm{V}}\left(V_{*}-b\right)}.
\end{equation}

Similarly,

\begin{equation}
\left(\frac{\partial P}{\partial V}\right)_{S_{\mathrm{th}}}=\frac{\partial}{\partial V}\left(\frac{RT}{V-b}-\frac{a}{V^{2}}\right)=-\frac{RT}{(V-b)^{2}}+\frac{R}{V-b}\left(\frac{\partial T}{\partial V}\right)_{S_{\mathrm{th}}}+\frac{2a}{V^{3}}.
\end{equation}

From the entropy relation, 
\begin{equation}
\left(\frac{\partial T}{\partial V}\right)_{S_{\mathrm{th}}}=-\frac{RT}{C_{\mathrm{V}}(V-b)}
\end{equation}

and therefore 
\begin{equation}
\left(\frac{\partial P}{\partial V}\right)_{S_{\mathrm{th}}}=-\frac{RT}{(V-b)^{2}}-\frac{R^{2}T}{C_{\mathrm{V}}\left(V-b\right)^{2}}+\frac{2a}{V^{3}}.
\end{equation}

Evaluating this expression at equilibrium gives

\begin{equation}
\left.\frac{\partial P}{\partial V}\right|_{S_{\mathrm{th}},*}=-\frac{RT_{\mathrm{b}}}{\left(V_{*}-b\right)^{2}}\left(1+\frac{R}{C_{\mathrm{V}}}\right)+\frac{2a}{V_{*}^{3}}.
\end{equation}

Substituting these results into the Hessian of $\left.\nabla^{2}S_{\mathrm{NB}}\right|_{*}$,
we obtain 
\begin{equation}
\left.\nabla^{2}S_{\mathrm{NB}}\right|_{*}=-\left(\begin{array}{ccc}
\frac{R}{\left(V_{*}-b\right)^{2}}\left(1+\frac{R}{C_{\mathrm{V}}}\right)-\frac{2a}{V_{*}^{3}T_{\mathrm{b}}} & 0 & -\frac{R}{C_{\mathrm{V}}\left(V_{*}-b\right)}\\
0 & \frac{1}{mT_{\mathrm{b}}} & 0\\
-\frac{R}{C_{\mathrm{V}}\left(V_{*}-b\right)} & 0 & \frac{1}{C_{\mathrm{V}}}
\end{array}\right).
\end{equation}

It follows that the covariance matrix is 
\begin{equation}
C=k_{\mathrm{B}}\left(\begin{array}{ccc}
\frac{1}{K_{*}} & 0 & \frac{R}{\left(V_{*}-b\right)K_{*}}\\
0 & mT_{\mathrm{b}} & 0\\
\frac{R}{\left(V_{*}-b\right)K_{*}} & 0 & C_{\mathrm{V}}+\frac{R^{2}}{\left(V_{*}-b\right)^{2}K_{*}}
\end{array}\right),
\end{equation}

where 
\begin{equation}
K_{*}=\frac{R}{\left(V_{*}-b\right)^{2}}-\frac{2a}{V_{*}^{3}T_{\mathrm{b}}}=-\frac{1}{T_{\mathrm{b}}}\left.\left(\frac{\partial P}{\partial V}\right)_{T}\right|_{*}.
\end{equation}

It is useful to note that 
\begin{equation}
K_{*}>0\Longleftrightarrow\nabla^{2}S_{\mathrm{NB}}<0\Longleftrightarrow\left(\frac{\partial P}{\partial V}\right)_{T}<0.
\end{equation}

Thus, the condition that $S_{\mathrm{NB}}$ have a local maximum reproduces
the usual thermodynamic stability condition.

The fluctuations are therefore given by 
\begin{equation}
\langle\left(\delta V\right)^{2}\rangle=\frac{k_{\mathrm{B}}}{K_{*}},
\end{equation}

\begin{equation}
\langle\left(\delta p\right)^{2}\rangle=mk_{\mathrm{B}}T_{\mathrm{b}},
\end{equation}

\begin{equation}
\langle\delta V\delta S_{\mathrm{th}}\rangle=\frac{k_{\mathrm{B}}R}{\left(V_{*}-b\right)K_{*}},
\end{equation}

and

\begin{equation}
\langle(\delta S_{\mathrm{th}})^{2}\rangle=k_{\mathrm{B}}\left[C_{\mathrm{V}}+\frac{R^{2}}{\left(V_{*}-b\right)^{2}K_{*}}\right].
\end{equation}

In the limit $a,b\to0$, these expressions reduce to those of the
ideal gas.

Furthermore, as 
\begin{equation}
K_{*}\to0^{+},
\end{equation}

we have

\begin{equation}
\langle\left(\delta V\right)^{2}\rangle\to\infty,
\end{equation}
\begin{equation}
\langle\delta V\delta S_{\mathrm{th}}\rangle\to\infty,
\end{equation}

and

\begin{equation}
\langle(\delta S_{\mathrm{th}})^{2}\rangle\to\infty.
\end{equation}

This corresponds to the spinodal condition 
\begin{equation}
\left(\frac{\partial P}{\partial V}\right)_{T}=0.
\end{equation}

Thus, for the van der Waals gas, the Hessian of $\ensuremath{S_{\mathrm{NB}}}$
becomes degenerate on the spinodal line, and this degeneracy determines
the divergence of the $V-S_{{\rm th}}$ fluctuations. In other words,
the curvature of $S_{\mathrm{NB}}$ becomes flat along one direction,
while the corresponding covariance, being determined by the inverse
Hessian, diverges. Thermodynamic instability can therefore be interpreted
as a flattening of $S_{\mathrm{NB}}$, manifested statistically through
divergent fluctuations.

At the critical point, one additionally has 
\begin{equation}
\left(\frac{\partial^{2}P}{\partial V^{2}}\right)_{T}=0,
\end{equation}

with 
\begin{equation}
V_{c}=3b,\ T_{c}=\frac{8a}{27Rb}.
\end{equation}

Since the Hessian is already degenerate on the spinodal line, it is
also degenerate at the critical point. Consequently, the Gaussian
saddle-point approximation ceases to be valid in the vicinity of criticality.

\subsection{Generalized forces $\beta_{a}^{\mathrm{NB}}=\partial s_{\mathrm{NB}}/\partial H_{a}$
and Nambu thermodynamics}

When the stationary distribution is given by 
\begin{equation}
\phi_{\mathrm{st}}=\frac{1}{Z_{\mathrm{NB}}}\exp\left(\frac{S_{\mathrm{NB}}(H_{1},H_{2})}{k_{\mathrm{B}}}\right),
\end{equation}

the Hamiltonians $H_{1}$ and $H_{2}$ serve not only as dynamical
invariants of the Nambu sector but also as constraint variables characterizing
the statistical ensemble underlying the stationary distribution. In
terms of the dimensionless potential 
\begin{equation}
s_{\mathrm{NB}}=\frac{S_{\mathrm{NB}}}{k_{\mathrm{B}}},
\end{equation}

we may define the generalized thermodynamic forces 
\begin{equation}
\beta_{a}^{\mathrm{NB}}\equiv\frac{\partial s_{\mathrm{NB}}}{\partial H_{a}}\qquad(a=1,2).
\end{equation}

The differential of $s_{\mathrm{NB}}$ can then be written as 
\begin{equation}
ds_{\mathrm{NB}}=\beta_{1}^{\mathrm{NB}}dH_{1}+\beta_{2}^{\mathrm{NB}}dH_{2}.
\end{equation}

When $H_{1}$ represents the energy, $\beta_{1}^{\mathrm{NB}}$ can
be interpreted as an inverse-temperature-like quantity. By contrast,
$H_{2}$ is an additional constraint associated with the Nambu structure,
and $\beta_{2}^{\mathrm{NB}}$ is the generalized force conjugate
to this constraint. In this sense, the Nambu canonical distribution
extends the ordinary canonical ensemble, which is organized by a single
energy constraint, to a nonequilibrium stationary ensemble characterized
by multiple Nambu Hamiltonians\footnote{A related construction was already suggested by Nambu in his original
paper. Under an ergodic assumption, he considered a generalized canonical
weight of the form $\rho\propto\exp\left(-\beta H-\gamma C\right)$
for a system with two conserved Hamiltonians and interpreted $\beta$
and $\gamma$ as two temperaturelike intensive parameters, in analogy
with a grand-canonical ensemble \cite{Nambu1973}. In the present
formulation, however, the conjugate parameters $\beta_{a}^{\mathrm{NB}}$
need not all represent physical inverse temperatures; more generally,
they are generalized intensive variables conjugate to the Nambu Hamiltonians.
The present result differs in that the exponential weight is derived
from the stationary Fokker--Planck equation of stochastic open-system
NNET, with $S_{\mathrm{NB}}$ identified as the stationary quasipotential.}.

\section{Discussion}

In Nambu nonequilibrium thermodynamics, the relation between the dissipation
potential $S_{\mathrm{NB}}$ and the self-information $S_{\mathrm{Sh}}$
in a nonequilibrium stationary state is given by 
\begin{equation}
S_{\mathrm{Sh}}(x)=\log Z_{\mathrm{NB}}-\frac{S_{\mathrm{NB}}(x)}{k_{\mathrm{B}}}
\end{equation}

This result should not be interpreted as identifying $S_{\mathrm{NB}}$
with the Shannon entropy itself. Rather, it shows that $S_{\mathrm{NB}}$
acts as the quasipotential that determines the stationary probability
distribution.

\subsection{Thermodynamic entropy and self-information are distinct}

The thermodynamic entropy $S_{\mathrm{th}}$ is a state function determined
by thermodynamic variables such as temperature, volume, and particle
number. By contrast, the self-information 
\begin{equation}
S_{\mathrm{Sh}}(x,t)=-\ln\phi(x,t)
\end{equation}

is an information-theoretic quantity determined by a probability distribution.
Even for the same macroscopic state, its value depends on the probability
distribution under consideration. The two quantities therefore do
not coincide in general.

\LyXZeroWidthSpace{}

In equilibrium statistical mechanics, Boltzmann's relation 
\begin{equation}
S=k_{\mathrm{B}}\log W
\end{equation}

connects thermodynamic entropy with the logarithm of the number of
accessible microscopic states. This correspondence, however, relies
on assumptions such as equilibrium, an appropriate coarse graining,
and the equivalence of statistical ensembles. In an open nonequilibrium
system, the thermodynamic entropy $S_{\mathrm{th}}$ of the subsystem
may vary nonmonotonically because of exchanges with thermal or particle
reservoirs. It therefore cannot, in general, be directly identified
with either the local self-information or the Shannon entropy.

\LyXZeroWidthSpace{}

The present results instead indicate that the three quantities

\begin{equation}
S_{\mathrm{th}},\qquad S_{\mathrm{NB}},\qquad S_{\mathrm{Sh}}
\end{equation}

should be distinguished. Here, $S_{\mathrm{th}}$ is the thermodynamic
state function, $S_{\mathrm{NB}}$ is the dissipation potential generating
the irreversible gradient flow, and $S_{\mathrm{Sh}}$ is the local
information measure determined by the stationary probability distribution.

\subsection{$S_{\mathrm{NB}}$ as a stationary quasipotential}

The relation derived in this work implies that $S_{\mathrm{NB}}$
acts as a stationary quasipotential in the sense that it determines
the logarithmic weight of the nonequilibrium stationary distribution.
In large-deviation theory, a quasipotential governs the exponential
asymptotics of the stationary distribution of a stochastic dynamical
system \cite{FreidlinWentzell_2012}. Similarly, the dissipation potential
obtained here determines the stationary probability weight through
\begin{equation}
\phi_{\mathrm{st}}(x)\propto\exp\left[\frac{S_{\mathrm{NB}}(x)}{k_{\mathrm{B}}}\right].
\end{equation}

It should be emphasized, however, that $S_{\mathrm{NB}}$ is not defined
here as the general Freidlin--Wentzell quasipotential. In the standard
large-deviation convention, the stationary density is commonly written
as $\phi_{\mathrm{st}}\asymp\exp[-V/\epsilon]$; thus, $S_{\mathrm{NB}}/k_{\mathrm{B}}$
corresponds, up to conventions and additive constants, to the negative
of the associated rate function. In the present formulation, $S_{\mathrm{NB}}$
is first introduced as a macroscopic potential generating the dissipative
gradient flow of NNET. The central result is that, under the conditions
identified above, the same potential also governs the stationary probability
distribution and thereby acquires the role of a stationary quasipotential.

This construction is related in spirit to the steady-state thermodynamics
of Sasa and Tasaki \cite{Sasa_2006}. Whereas their framework defines
thermodynamic quantities through operational procedures for nonequilibrium
steady states, the present formulation starts from stochastic Nambu
dynamics and identifies $S_{\mathrm{NB}}$ as the stationary quasipotential.
Whether this potential can also be related to an operational nonequilibrium
entropy through an extended Clausius relation remains to be clarified.

\subsection{Generalized thermostatistical Nambu systems and q-generalized NNET}

In the present work, we have restricted the stationary distribution
to the ordinary exponential form

\begin{equation}
p_{\mathrm{NB}}(x)\propto\exp\left[s_{\mathrm{NB}}(x)\right].
\end{equation}

Within this setting, the saddle-point approximation around an isolated
nondegenerate maximum that is also a fixed point of the Nambu flow
yields ordinary Gaussian fluctuations. When strong nonlinearities,
long-range correlations, or non-Gaussian fluctuations become important,
however, a generalization based on $q$-exponential distributions
provides a natural possibility \cite{Tsallis_1988}.

\LyXZeroWidthSpace{}

In this direction, $q$-logarithmic likelihoods, $q$-exponential
distributions, and $q$-Massieu functions can be used to generalize
ensemble equivalence and the Gaussian principle \cite{Wada2013Gauss,Wada_2024}.
Importantly, Frank has already developed a generalized thermostatistical
formulation of stochastic active and purely dissipative Nambu systems.
His framework incorporates generalized entropy measures, including
the Tsallis entropy, through nonlinear free-energy Fokker--Planck
equations and derives stationary probability densities when the dissipative
or pumping mechanism takes a canonical-dissipative form and depends
only on the Nambu invariants \cite{Frank_2016}.

\LyXZeroWidthSpace{}

Accordingly, the next step is not merely to replace the ordinary exponential
weight of NNET by a $q$-exponential one. Rather, it is to reinterpret
Frank's generalized thermostatistical construction within the framework
of Nambu nonequilibrium thermodynamics. In particular, it will be
necessary to determine whether a $q$-deformed counterpart of $S_{\mathrm{NB}}$
can be identified as an open-system dissipation potential generating
the irreversible sector, and whether the corresponding stationary
distribution can be derived directly from a $q$-generalized stochastic
NNET equation. Such an analysis should also clarify the relations
among the $q$-generalized self-information, generalized statistical
entropy, and the Massieu-type dissipation potential, as well as the
conditions under which the Nambu flow preserves the latter.

\LyXZeroWidthSpace{}

This program would connect Frank's generalized-entropy description
of stochastic Nambu systems with the thermodynamic distinction among
\begin{equation}
S_{\mathrm{th}},\qquad S_{\mathrm{NB}},\qquad S_{\mathrm{Sh}}
\end{equation}

developed in the present work.

\subsection{Limitations in strongly nonlinear systems and the global existence
of $S_{\mathrm{NB}}$}

The analysis presented in this paper relies on several conditions.
First, the dissipative sector must be expressible as a gradient flow
generated by a single scalar potential $S_{\mathrm{NB}}$. Second,
an Einstein-type relation must hold between the diffusion matrix and
the transport tensor. Third, the Nambu flow must preserve $S_{\mathrm{NB}}$,
namely,

\begin{equation}
N\cdot\nabla S_{\mathrm{NB}}=0.
\end{equation}

When any of these conditions is violated, the stationary distribution
need not take the simple form

\begin{equation}
\phi_{\mathrm{st}}\propto\exp\left(\frac{S_{\mathrm{NB}}}{k_{\mathrm{B}}}\right).
\end{equation}

In particular, for strongly nonlinear or chaotic systems, a globally
smooth dissipation potential $S_{\mathrm{NB}}$ may not exist. This
issue is closely related to the local character of the Helmholtz--Darboux
decomposition underlying NNET and to the obstructions encountered
when attempting to extend it globally \cite{katagiri2025nambu2}.
In the absence of a global dissipation potential, the logarithmic
weight of the stationary distribution may have to be described in
terms of quasipotentials defined on local patches, nonconservative
probability currents, or the curvature of a nonintegrable thermodynamic
connection rather than by a single global scalar function.

\LyXZeroWidthSpace{}

This perspective is naturally connected to the gauge-fixing formulation
of nonequilibrium thermodynamics. In that formulation, the thermodynamic
force is treated as a one-form, 
\begin{equation}
A=A_{i}da^{i}.
\end{equation}

When this one-form is integrable, it can locally be written as

\begin{equation}
A=dS,\ F=dA=0.
\end{equation}

For a generalized thermodynamic connection, however,

\begin{equation}
F=dA\neq0
\end{equation}

is allowed. A nonvanishing curvature expresses the impossibility of
representing the thermodynamic force by a single globally defined
entropy potential and therefore characterizes nonintegrability and
path dependence \cite{katagiri_gauge_fixing}.

\LyXZeroWidthSpace{}

Furthermore, in the gauge-theoretic formulation of NNET, the Nambu
bracket emerges from a coupling of the form

\begin{equation}
A\wedge B\wedge dt,
\end{equation}

together with a local Darboux--Clebsch representation of a closed
two-form,

\begin{equation}
B=dH_{1}\wedge dH_{2}.
\end{equation}

Although this construction reproduces the NNET evolution law locally,
the global patching of $A$ and $B$, the possible nonintegrability
of the thermodynamic connection, and the presence of curvature provide
a natural geometric setting for situations in which no single global
$S_{\mathrm{NB}}$ exists \cite{katagiri_gauge_NNET}. The limitations
of a global $S_{\mathrm{NB}}$ should therefore be understood not
as a breakdown of NNET itself, but as a gauge-geometric problem involving
the patchwise definition and gluing of local potentials.

\LyXZeroWidthSpace{}

Accordingly, the present results do not imply that $S_{\mathrm{NB}}$
increases monotonically or universally determines the stationary distribution
in every nonequilibrium system. Rather, our claim is conditional:
when the nondissipative Nambu flow and the irreversible gradient flow
satisfy the compatibility conditions stated above, $S_{\mathrm{NB}}$
serves simultaneously as the dissipation potential and as the stationary
quasipotential governing the probability distribution. Clarifying
this conditional structure provides the starting point for a statistical
interpretation of stochastic NNET.

\section*{Acknowledgments}

S.K. would like to thank Akio Sugamoto and Yoshiki Matsuoka for their
valuable discussions and collaboration throughout the development
of Nambu nonequilibrium thermodynamics. He is also grateful to Shiro
Komata for carefully reading the manuscript and providing helpful
comments. Finally, S.K. thanks the members of the Field Theory Seminar
at The Open University of Japan (OUJ) for providing a stimulating
forum in which the NNET program could develop.

 \bibliographystyle{unsrt}
\bibliography{nnet_diff_new_v2}

\end{document}